\documentclass{aa}

\usepackage[colorlinks=true, linkcolor=blue, citecolor=blue, urlcolor=blue]{hyperref}
\usepackage{graphicx}
\usepackage{verbatim}
\usepackage{soul}
\usepackage{threeparttable}
\usepackage[normalem]{ulem}

\usepackage{txfonts}

    \newcommand{\nustar}{\textit{NuSTAR}}
    
    \newcommand{\rxte}{\textit{RXTE}}
    \newcommand{\axsini}{$a_{x}$sin(\textit{i})}
    \newcommand{\funi}{erg cm$^{-2}$ s$^{-1}$}
\begin{document} 

   \title{Dynamical constraints on the companion mass from pulsar timing and cyclotron line confirmation in the SFXT IGR J16418--4532}
 \titlerunning{SFXT IGR J16418--4532}

   \author{Ashwin Devaraj\inst{1,2} \fnmsep\thanks{E-mail: ashwin.d@iitb.ac.in},
          Rahul Sharma\inst{3,1},
          Ketan Rikame\inst{1},
          \and
          Biswajit Paul\inst{1}%
          }
\authorrunning{Devaraj et al.}

   \institute{Raman Research Institute, C V Raman Avenue, Sadashivanagar, Bangalore 560080, India%
        \and 
            Indian Institute of Technology Bombay, Powai, Mumbai 400076, India           
        \and
             Inter-University Centre for Astronomy and Astrophysics (IUCAA), Ganeshkhind, Pune 411007, India
          }

\abstract
{Supergiant Fast X-ray Transients (SFXTs) exhibit extreme X-ray variability on short timescales of minutes to hours, driven by accretion processes that are still poorly understood. IGR J16418–4532 is an intermediate SFXT in which X-ray pulsations have been detected and which has been proposed as a candidate cyclotron line source, making it a key system for probing neutron-star accretion physics and magnetic field properties in this class of sources.}
{We aim to constrain the projected semi-major axis of the SFXT IGR J16418--4532, through timing analysis, and confirm the presence of a Cyclotron Resonant Scattering Feature (CRSF), thereby characterising the nature of the compact object in this system.}
{We analysed a 148~ks long \textit{NuSTAR} observation and several archival \textit{RXTE} observations using epoch-folding techniques to search for pulsations while correcting the photon arrival times for a range of \axsini \ values. We performed phase-averaged spectroscopy using several different continuum models and assessed the significance of the cyclotron line feature.}
{We detect coherent pulsations with $P_{\mathrm{spin}} = 1208.67 \pm 0.04$~s and $P_{\mathrm{spin}} = 1209.4178 \pm 0.0084$~s in \textit{NuSTAR} and \textit{RXTE/PCA} light curves, respectively. The pulsations exhibit clear dependence on the assumed projected semi-major axis, and are detected with the highest significance for $a_x\sin i = 55.1 \pm 3.5$~lt-s (\textit{NuSTAR}) and $51.3 \pm 2.6$~lt-s (\textit{RXTE/PCA}). The light curves show several flares, and the pulse profiles are found to evolve over the course of the observation. Sliding window pulse profiles reveal the pronounced effect of the orbital delay on the pulse arrival times in uncorrected data, which is fully removed post-orbital correction. Using the inferred orbital parameters, we constrain the mass and radius of the companion star to be in the range of $\sim$12--18 $M_\odot$ and $\sim$13--16 $R_\odot$, respectively. We report the first successful constraint on the binary mass function of an SFXT obtained by considering the orbital motion of the neutron star. Spectral analysis reveals a robust cyclotron absorption line at $\sim$50~keV, independent of the choice of the continuum, suggesting a magnetic field strength of $\sim$$5 \times 10^{12}$~G.}
{The cyclotron line confirmation and dynamical constraints on the companion's mass indicate that IGR J16418--4532 hosts a neutron star with a magnetic field of B$\approx 5 \times 10^{12}~G$ accreting from a $\sim$12--18 $M_\odot$ companion. We demonstrate that optimising the pulse detection by correcting the photon arrival times with trial \axsini \ values rather than modelling the pulse arrival time delays offers a route to obtain dynamical mass measurements in other SFXTs whose variable pulse profiles limit the scope of conventional timing techniques.}


   \maketitle

\section{Introduction}
Supergiant Fast X-ray Transients (SFXTs) are a subclass of High Mass X-ray Binaries (HMXBs) in which the compact object accretes matter from a supergiant OB stellar companion, showing extreme transient behaviour. In contrast to classical Supergiant X-ray Binaries (SgXBs), which exhibit persistent or slowly varying X-ray emission, SFXTs show rapid variability on short time scales, ranging from a few minutes to hours. However, there are some extremely flaring SgHMXBs, such as LMC X-4 \citep{Levine_2000, Beri_2017}, EXO 2030+375 \citep{Parmar_1989}, and SMC X-1 \citep{Moon_2003}; the flares in these systems are relatively rare. A few of the SFXTs span a large dynamic range from 10$^{32}~\mathrm{erg~s^{-1}}$ in quiescence to about $10^{38}~ \mathrm{erg~s^{-1}}$ at the peak of their flares \citep{Sidoli_2017}. For reviews, see \citet{Sguera_2006}, \citet{Sidoli_2013}, \citet{Walter_2015}, \citet{Negueruela_2019}, \citet{Kretschmar_2021}. Recent multi-wavelength investigations have further explored the differences between SFXTs and persistent supergiant X-ray binaries \citet{van_den_Eijnden_2025}.

 A dozen SFXTs have been discovered so far \citep{Ducci_2014}, with the vast majority identified in a Galactic plane monitoring survey with \textit{INTEGRAL} \citep{Sguera_2005}. The orbital periods of these systems range from a few days to approximately 164 days, with a median of around 26 days \citep{Sidoli_2018}. IGR J16479--4514 and IGR J16418--4532 are two known eclipsing SFXTs with orbital periods of around 3.3~days \citep{Jain_2009} and 3.7~days \citep{Corbet_2006}, respectively. The measured eccentricities of some of these systems are < 0.4 \citep{Sidoli_2018}, while a couple of them, such as \mbox{IGR J08408--4503 \citep{Gamen_2015}} and IGR J11215--5952 \citep{Romano_2007}, with long orbital periods (>9 days), are expected to have eccentricities >0.6. In comparison to SgXRBs, SFXTs have been found to exhibit systematically lower absorption, average luminosities, and weaker fluorescence K$\alpha$ lines \citep{Pradhan_2018}.

Pulsations have been confirmed in very few of them, including the source of interest in this work, IGR J16418--4532 \citep{Walter_2006, Sidoli_2012}. Even in cases where pulsations have not been detected, the spectral shapes of these systems are generally well described by a power-law modified by an exponential cutoff between 10--30 keV \citep{Ducci_2010, Sidoli_2009, Romano_2011}, as is typically the case in accreting pulsars. The discovery of a cyclotron line at 17 keV in an SFXT was first reported by \citet{Bhalerao_2015} in IGR J17544--2619, confirming that this system hosted a highly magnetised neutron star despite the non-detection of pulsations. IGR J18483--0311 \citep{Sguera_2010} and IGR J16418--4532 \citep{Natalucci_2024} have been proposed to be candidate cyclotron line sources with lines tentatively present at 3.3~keV and 50~keV, respectively. Despite overwhelming evidence for the presence of a neutron star in these systems, the nature of the compact object in SFXTs remains debated because of the lack of clear pulsations and the detection of stable cyclotron lines in these sources. 

The characteristic behavior of SFXTs, involving rapid flares followed by long periods of quiescence has not been explained unambiguously yet. Several models that have been proposed involve either the presence of clumpy winds from the companion star \citep{Walter_2008} or being a result of gating mechanisms, such as a centrifugal barrier \citep{Grebenev_2010} or a magnetic barrier \citep{Bozzo_2008}, owing to the compact object being a highly magnetised neutron star. An alternative explanation for the flaring activity in SFXTs was provided by the Quasi-spherical settling accretion model \citep{Shakura_2012,Shakura_2015}. In this framework, hot plasma forms a quasi-static shell above the neutron-star magnetosphere, while inefficient cooling limits the rate at which plasma enters the magnetosphere, thereby suppressing the accretion rate onto the neutron star. An instability, potentially triggered by magnetic reconnection with magnetised stellar-wind material, can enhance plasma entry through the magnetosphere. The resulting increase in X-ray luminosity leads to efficient Compton cooling of the surrounding plasma, increasing its infall velocity and potentially causing the entire shell to accrete onto the neutron star on approximately the free-fall timescale, producing a bright X-ray flare. 

\begin{figure*}
    \centering
 \vspace{0.5cm}
	\includegraphics[scale=1.0,trim={0cm 0cm 0 0},  angle=0]{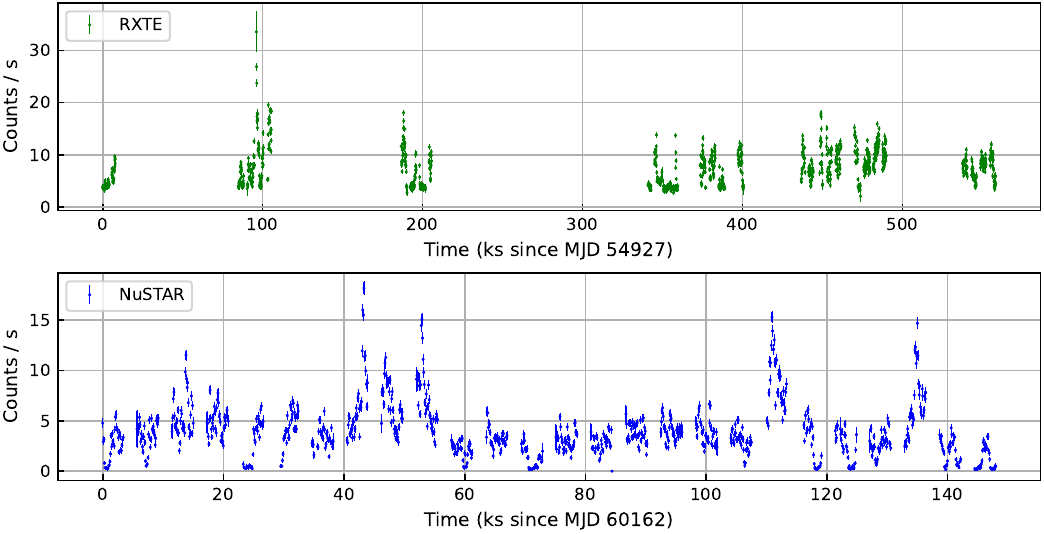}
    \caption{\textit{RXTE} light curve (top) and \textit{NuSTAR} light curve (bottom) binned at 100~s in the 2-16~keV and 3-79~keV range, respectively.}
    \label{fig:nus-rxte-ltc}
\end{figure*}

\begin{figure*}
    \centering
 \vspace{0.5cm}
	\includegraphics[scale=0.45,trim={0cm 0cm 0 0},  angle=0]{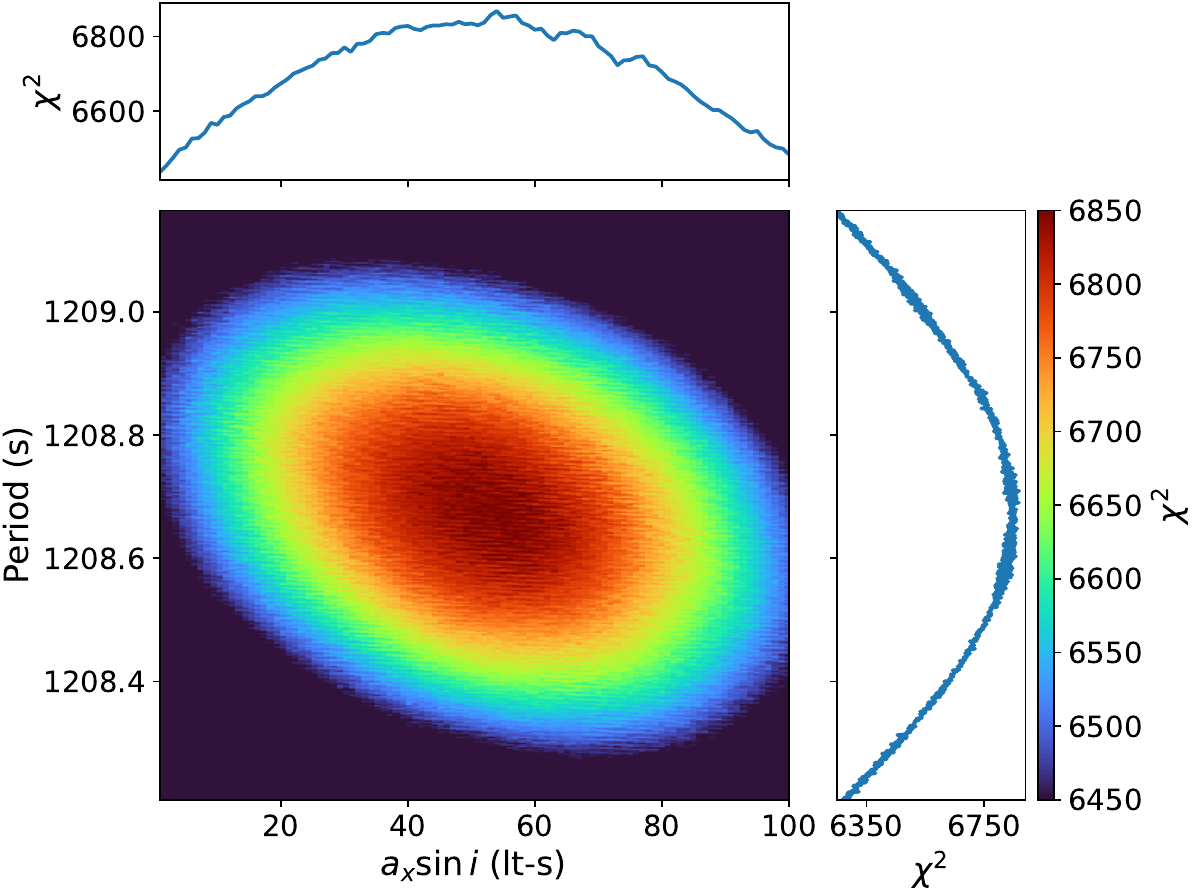}
    \includegraphics[scale=0.45,trim={0cm 0cm 0 0},  angle=0]{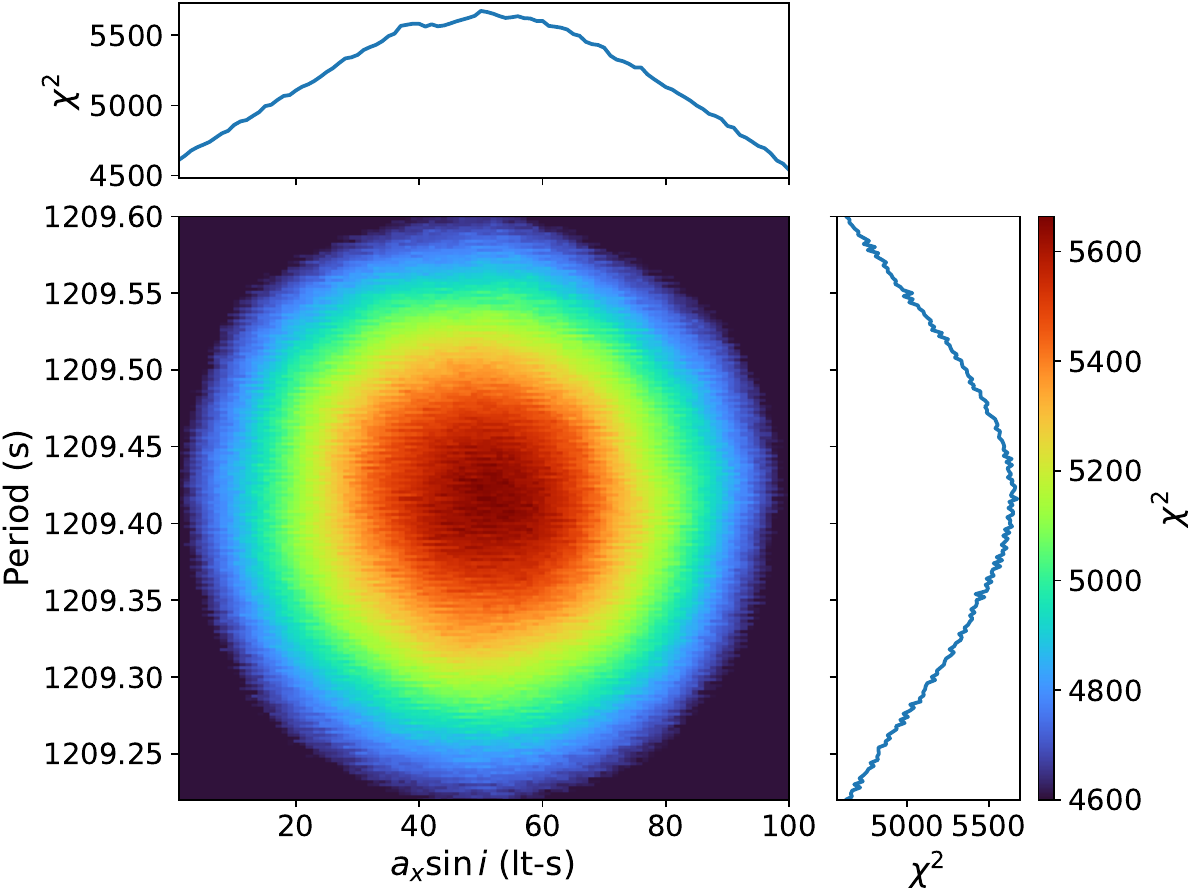}
    \caption{Two–dimensional epoch–folding $\chi^2$ statistic as a function of spin period $P_{\rm spin}$ and projected semi-major axis $a_x \sin i$, derived from NuSTAR (left) and RXTE (right) light curves. The colour scale represents the pulse detection significance, with the maximum indicating the best-fit parameter combination. The panels above and to the right show the corresponding one–dimensional $\chi^2$ profiles, obtained by maximising over the other parameter. For both light curves, the maximum $\chi^2$ is seen when the \axsini\ is approximately  50--55~lt-s.}
    \label{fig:efs-axsini-period-nus-rxte}
\end{figure*}
\begin{figure}
    \centering
 \vspace{0.5cm}
	\includegraphics[scale=0.6,trim={1cm 0cm 0 0},  angle=0]{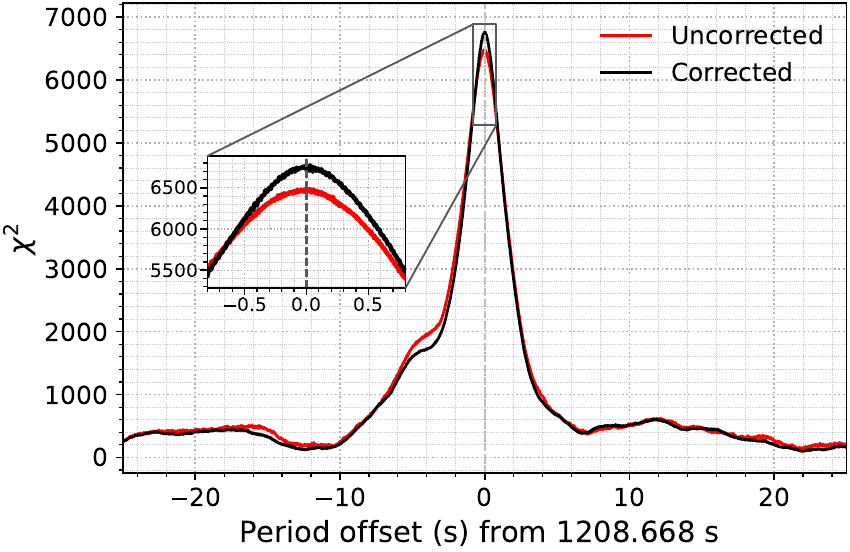}
    \includegraphics[scale=0.6,trim={1cm 0cm 0 0},  angle=0]{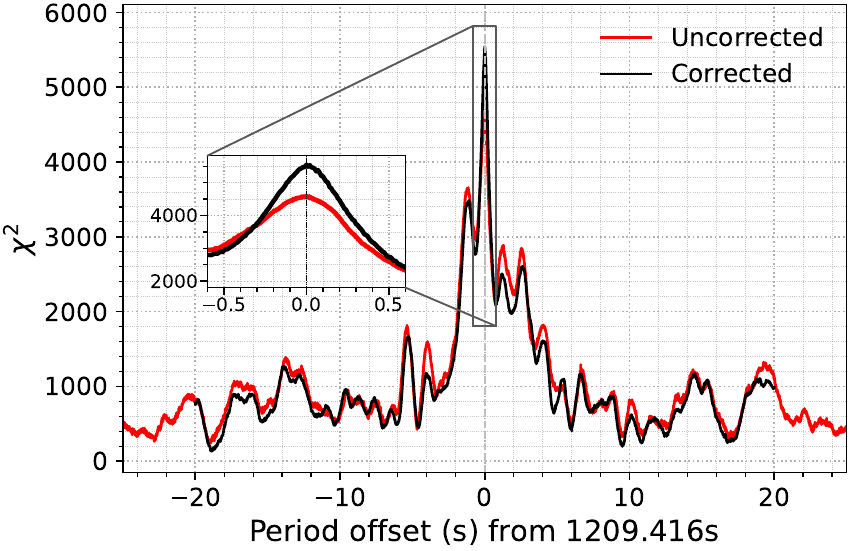}
    \caption{Comparison of the results of the epoch folding and $\chi^2$ maximisation for different periods with and without orbital correction. Top: \textit{NuSTAR} and Bottom: \textit{RXTE}.}
    \label{fig:efs-nus-rxte}
\end{figure}

 The discovery of the SFXT, IGR J16418--4532, was first reported following \textit{INTEGRAL} observations in  February 2003 by \citet{Tomsick_2004}. Follow-up monitoring with \textit{INTEGRAL} revealed recurrent, short‐duration X-ray flares, prompting \citet{Sguera_2006} to identify it as a candidate SFXT. Subsequent timing analysis of \rxte/ASM and \textit{Swift}/BAT light curves by \citet{Corbet_2006} yielded an orbital period of $\sim \!3.74$~d and revealed the presence of an eclipse in this system. \citet{Chaty_2008} modelled the optical/NIR SED and concluded that the companion is an OB-type supergiant with a minimal distance of around 13~kpc. The nature of the companion star is still under debate. \citet{Coleiro_2013} proposed a BN0.5Ia type companion based on the detection of He $\mathrm{I}$  lines in the optical/IR spectra. \citet{Goldoni_2012} suggest the companion's type to be an O9.5 due to the presence of strong He I lines in the K-band spectra. However, \citet{Coley_2015} argue that the companion must be a star of spectral type O7.5~$\mathrm{I}$ based on the duration of the eclipse ($\sim \!17\%  $ of the orbit) and the criteria for Roche-lobe accretion. \citet{Walter_2006} reported the discovery of pulsations in this source at $\sim1246\pm100$~s by analysing a 10~ks XMM-Newton observation. They detected several brief flares whose fluxes spanned roughly an order of magnitude with a heavily absorbed soft X-ray spectrum ($N_\mathrm{H} \sim 10^{23} \mathrm{{cm}}^{-2}$). \citet{Pradhan_2018} and \citet{Islam_2023} reported a moderate absorption column density of $\sim 5 \times 10^{22} ~\mathrm{cm}^{-2}$ with a low equivalent width of the Fe K$\alpha$. Similar to many SgXBs, a strong iron line was detected during eclipses \citep{Aftab_2019}. \citet{Romano_2012}, using \textit{Swift/XRT} data covering around 3 orbital periods, explain the X-ray emission from this source by applying the clumpy wind model and categorise this source as an intermediate SFXT.

Further evidence of the SFXT nature was reported by \citet{Sidoli_2012}. Using data from \textit{XMM-Newton}, they found the X-ray flux to vary by two orders of magnitude, with several bright X-ray flares of varying durations, from a few hundred to a thousand seconds. Based on observations, they suggest a transition in modes of accretion between pure wind accretion and Roche-lobe overflow. They also refined the spin period measurements to $ 1213 \pm 2$~s using the 2004 observation by \textit{XMM-Newton} and to $1209.4 \pm 1$~s using 2011 \textit{RXTE} observations and concluded the absence of significant spin up/spin down trends. The first detection of an X-ray dip, or ``off-state'', in an SFXT was reported for IGR J16418--4532 by \citet{Drave_2013}. The off-state was interpreted in the framework of quasi-spherical settling accretion as a transition from the Compton-cooling-dominated regime to the less efficient radiative-cooling regime \citep{Shakura_2013}. \citet{Poliakov_2017} investigated the flaring behaviour of this source  and argued that its long spin period favors quasi-spherical or magnetically controlled non-Keplerian accretion over the standard Keplerian accretion. This system is also known to exhibit a super-orbital modulation with a period $\sim 14.7$~days \citep{Corbet_2013, Islam_2019}. The strength of this modulation is found to vary on the timescale of a few years.

The detection of pulsations in most SFXTs remains tentative. Owing to large intensity fluctuations and associated changes in pulse profiles, dynamical mass measurements of the companion star have not yet been possible. The detection of CRSFs in these sources is also tenuous. In this work, we address two aspects of this binary, namely the detection of pulsations and the estimation of the mass of the companion through dynamical mass measurement, as well as the confirmation of the presence of a cyclotron line in this candidate source.

\section{Observations and Data Reduction}

The \textit{Nuclear Spectroscopic Telescope Array} (\nustar) is a hard X-ray telescope covering the 3–79 keV energy range \citep{Harrison_2013}. It focuses X-rays using grazing incidence optics onto two identical focal plane modules, FPMA and FPMB. Each module contains four CdZnTe detectors. The instrument provides an energy resolution of about 0.4 keV at 10 keV and around 0.9 keV at 60 keV. 

The \nustar\ observation of IGR J16418--4532 used in this work was performed on 6 August 2023 (OBSID:30901022002). This observation was 148~ks long, covering almost half the orbit of the binary, with an \mbox{exposure of $\sim 80$~ks}. The data were processed using the standard analysis suite \texttt{HEASOFT} v6.33c and \texttt{nustardas} v2.1.2 (CALDB version: 20231003).
Circular regions of size 75$^{\prime \prime}$ were chosen for the source and background regions on the same detector. The event files were barycenter-corrected using the tool \texttt{barycorr} and then, using the \texttt{nuproducts} script, we extracted the spectrum and light curve with a bin size of 1~s.  

IGR J16418--4532 was observed with the Rossi X-Ray Timing Explorer \citep[RXTE:][]{Bradt93} from 06 --16 April 2009 (observation ID: 94324-01). For the current work, we have used data from the observation made with the Proportional Counter Array \citep[PCA:][]{Jahoda96}. We have analysed data collected in the Good Xenon mode, which has a time resolution of 1 $\mu$s. The combined 2--16~keV light curve was extracted using the \textsc{seextract} tool. The background light curve was extracted from the Standard--2 mode data by using \textsc{pcabackest} with a faint source background model (pca\_bkgd\_cmfaintl7\_eMv20051128). Since a different number of PCUs were operational during the RXTE observations, we used the \textsc{correctlc} tool to convert the observed count rate to that of a single, mean PCU. The time series was barycenter corrected by using \textsc{faxbary}. For both \textit{NuSTAR} and \textit{RXTE} data, we used the entire observation.

IGR J16418--4532 has been continuously monitored with the \emph{Swift}/BAT hard X-ray transient monitor \citep{Krimm2013} since 2005. We used the publicly available $\sim$20-yr long (MJD 53415--60999) \emph{Swift}/BAT orbital light curve in the 15--50 keV energy band. The light curve was screened to remove data points of poor quality, retaining only those for which the data quality flag (\texttt{DATA\_FLAG}) was set to 0. The photon arrival times were subsequently corrected to the solar barycentre reference time using the \textsc{earth2sun} task.

\begin{figure}
    \centering
 \vspace{0.5cm}
    \includegraphics[scale=0.48,trim={1.5cm 0cm 0 0},  angle=0]{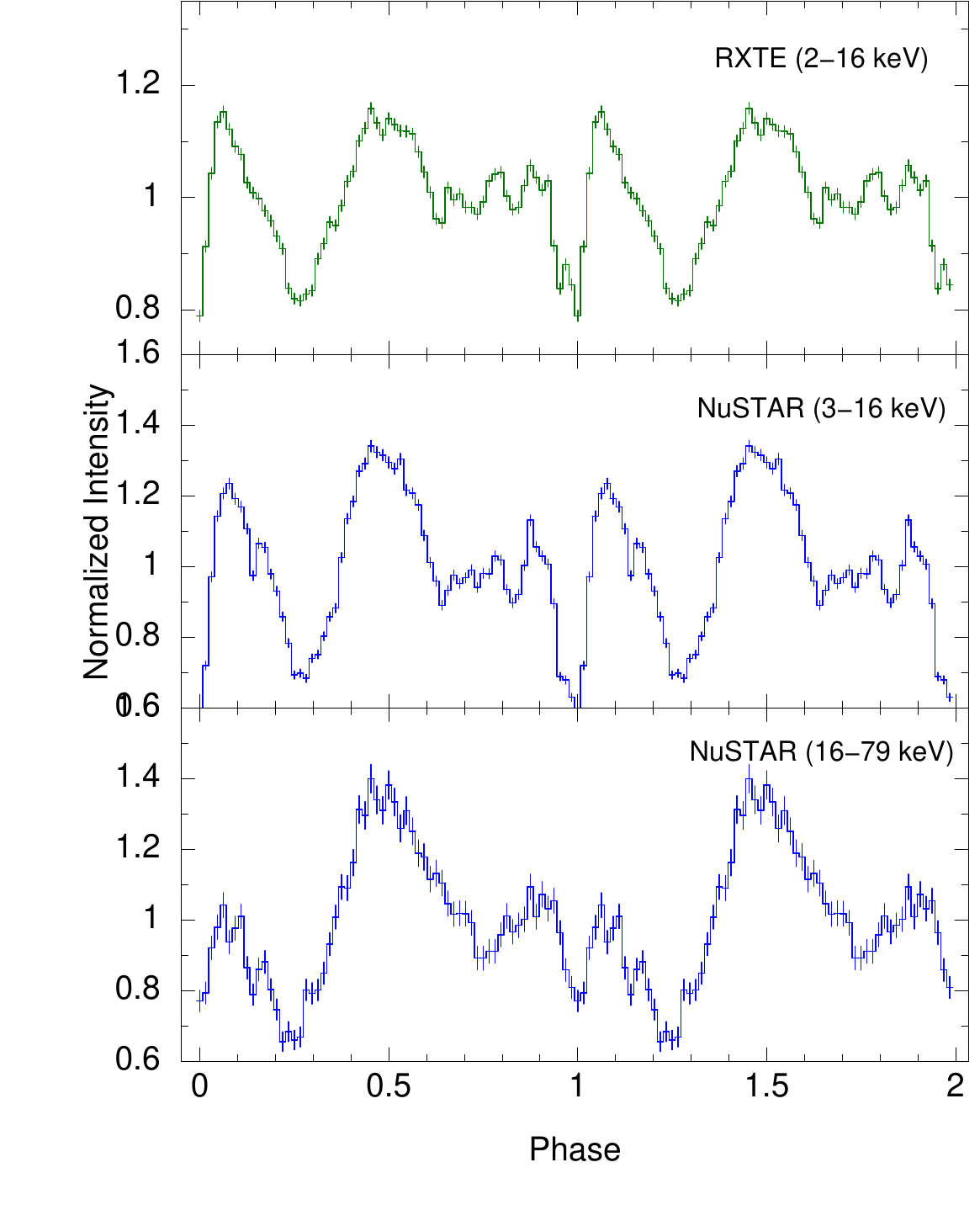}
    \caption{Pulse profile constructed using the entire light curves corrected for orbital motion. The \textit{RXTE} pulse profile (top) in the 2--16~keV range and the \textit{NuSTAR} pulse profiles in the 3--16~keV and 16--79~keV ranges respectively.}
    \label{fig:pprofs-nus-rxte}
\end{figure}

\begin{figure}
    \centering
 \vspace{0.5cm}
	\includegraphics[scale=1.0,trim={0cm 0cm 0 0}, ]{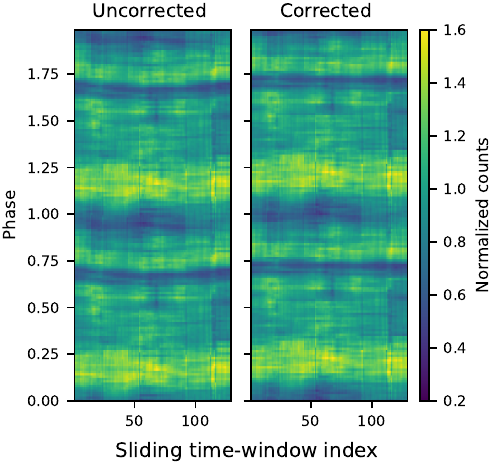}
    \caption{ Sliding-window pulse profiles constructed from the \textit{NuSTAR} light curve before (left) and after (right) applying orbital correction. Pulse profiles were generated using a fixed window of 20~ks, shifted in steps of 1~ks across the full 148~ks observation. The uncorrected profiles exhibit a systematic drift of the pulse minimum (around phase 0.7) with time, manifesting as a curved structure in phase, resulting from %
    orbital motion. After correcting photon arrival times for orbital effects, the pulse minima remain aligned at a constant phase across all sliding windows.} 
    \label{fig:sliding-pprofs-nus}
\end{figure}

\begin{figure*}
    \centering
 \vspace{0.5cm}
	\includegraphics[scale=0.45,trim={0cm 0cm 0 0},  angle=0]{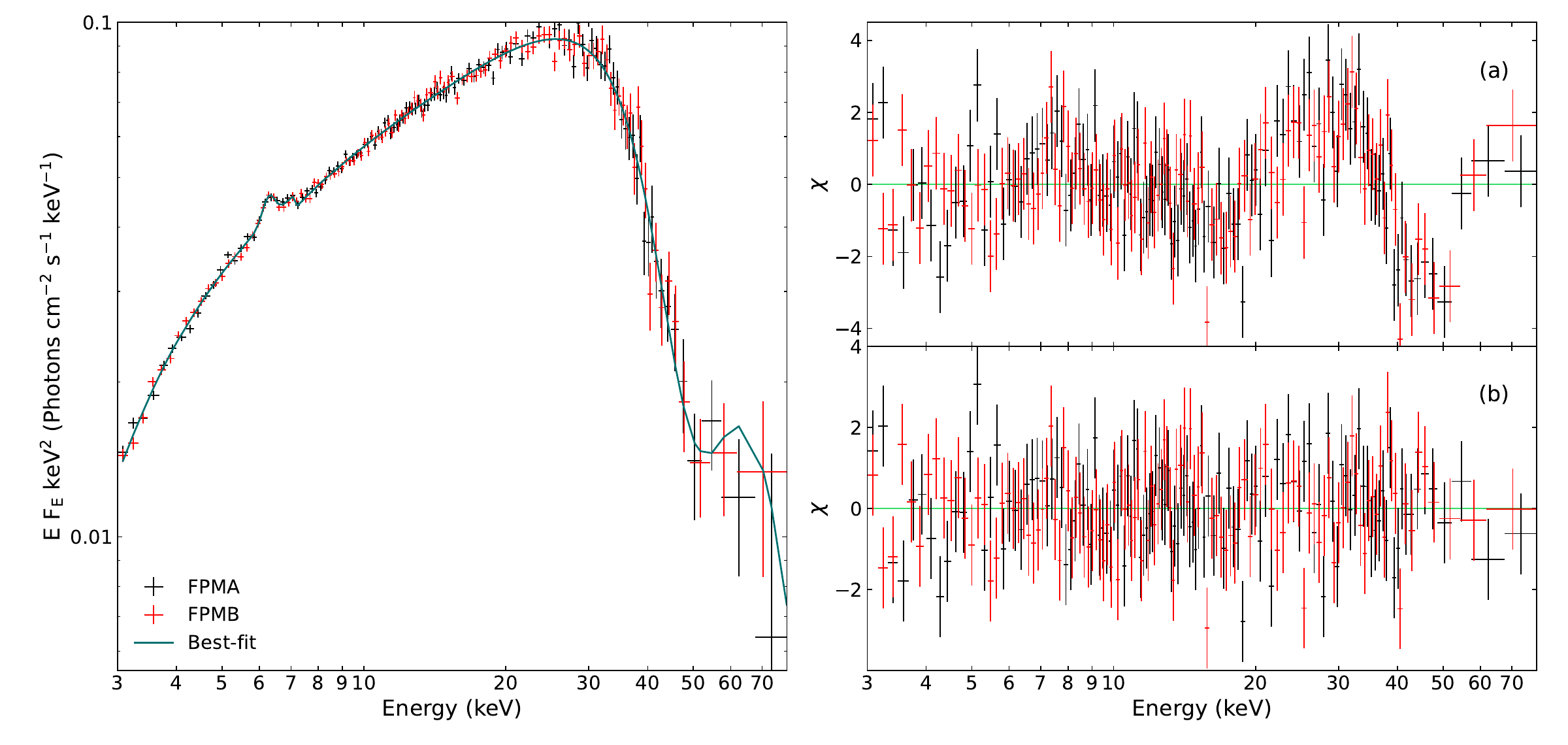}
    \caption{The unfolded spectrum with the best-fit of \texttt{NPEX} model is shown in the left panel. The right panel shows the residuals. Panel (a) is the fit without the cyclotron line at 50~keV, (b) is the best fit including a \texttt{gabs} at 50~keV. The red and black points correspond to data from FPMA and FPMB, respectively.}
    \label{fig:nustar-spectrum}
\end{figure*}

\section{Analysis and Results}

\subsection{Timing Analysis}
The \nustar\ and \rxte \ light curves are presented in Fig. \ref{fig:nus-rxte-ltc}.  We performed a period search on the 1~s light curves using the epoch-folding and $\chi^2$ maximisation technique \citep{Leahy_1987} using the \texttt{XRONOS} tool \texttt{efsearch}. The size of the binary system was not previously determined. The standard method for determining the orbital parameters of systems accreting X-ray pulsars is by modelling the pulse arrival time delay as a function of the binary orbital phase \citep{Raichur_2010}. This requires the source to have a stable pulse profile with time, which is not the case for IGR J16418--4532, as will be clear from subsequent analyses. Moreover, the orbital pulse arrival time delay technique provides accurate values for the orbital parameters when the pulse period is significantly smaller than the projected light travel time across the binary along the line of sight to the system, as seen in Cen X-3.

The orbital motion of the compact object can smear out the pulses if the appropriate Time of Arrival (ToA) corrections are not applied, even for a slow pulsar like this one. In order to probe the orbital dependence of the ephemeris of this pulsar, we corrected the light curves for a range of \axsini \ values from 1~lt-s to 100~lt-s with the following assumptions about the binary parameters. We assumed a circular orbit ($e=0$) since binary systems with short orbital periods ($<5$~days) tend to have low eccentricities \citep{Leiner_2015}. We used the orbital period and mid-eclipse time of P$_{\text{orb}} = 3.73881$~d and T$_{90}$=54930.694~MJD. An epoch-folding search yielded a best-fit orbital period of 3.73881\,(2)~d, consistent with \citet{Coley_2015}. The folded profile was fitted with a step-and-ramp function to model the eclipse, from which we determined the mid-eclipse epoch of MJD 54930.694\,(11) and eclipse duration of $\Delta T_{\rm ecl}=0.180\, (6)\, P_{\rm orb}$. We used the above-mentioned values of $T_{90}$ and $P_{\rm orb}$ for the pulsation search. 

We performed the epoch-folding period search on each of the hundred light curves and produced the heatmaps in Fig.\ref{fig:efs-axsini-period-nus-rxte}. For both \nustar\ and \rxte\ light curves, a clear dependence of the period on the \axsini \ value can be seen, where the $\chi^2$ peaks around 55 lt-s. In both cases, the \axsini\ correction significantly improves the detection. For \nustar, we find an improvement in $\chi^2$ of about 370 upon correcting the light curves, as shown in the top panel of Fig. \ref{fig:efs-nus-rxte}. We determine the spin period to be $P_{\mathrm{spin}} = 1208.67 \pm 0.04$~s and the \axsini \ = $55.1 \pm 3.5$~lt-s. 
For \textit{RXTE}, the improvement in $\chi^2$ was around 1080 upon correcting the light curves for the orbital motion, as shown in the bottom panel of Fig. \ref{fig:efs-nus-rxte}. We determine the period to be $P_{\mathrm{spin}} = 1209.4178 \pm 0.0084$~s and the \axsini \ = $51.3 \pm 2.6$~lt-s. 

The uncertainty on both these parameters was determined using the bootstrap method. We simulated 1,000 light curves using the original light curve, following the procedure outlined in \citet{Boldin_2013}. Following this, we produced a set of orbital-corrected light curves from each simulated light curve and performed an epoch-folding search. We then proceeded to identify the $P_{\mathrm{spin}}$ and \axsini\ values corresponding to the maximum $\chi^2$. The uncertainties quoted above are the standard deviations of all the periods and \axsini\ values.

\subsubsection{Pulse profiles}

Using the orbital-corrected light curves from \nustar \ and \rxte \  and the appropriate period, we constructed the pulse profiles, which are shown in Fig. \ref{fig:pprofs-nus-rxte}. The pulse profiles were constructed using the full exposure without excluding the flares as the count rates were found to vary only by a small factor unlike the extreme dynamic range seen in typical SFXTs. The pulse profiles primarily show two major peaks, around 0.1 and 0.5, and a third peak between phases 0.6 and 0.9, along with smaller structures. From Fig. \ref{fig:pprofs-nus-rxte}, it is apparent that the soft band pulse profiles of \textit{RXTE} and \textit{NuSTAR} are largely similar while the hard band shows a change in the peak around 0.1. Although minor energy-dependent variations can be seen, the morphology of the pulse profiles in the different bands remains broadly consistent. We also examined the flux dependence of the pulse profiles by folding the \nustar\ light curves for count rates below and above 10~cts~s$^{-1}$ and found the profiles to be broadly consistent except for minor differences near the second peak. As this observation samples a modest dynamic range, a study of the flux dependence of the pulse profiles is not feasible with the current data and would be better suited for future observations sampling a wider range of flux states.

To visualise the effect of orbital motion on the observed pulse phase, we constructed sliding-window pulse profiles using the \textit{NuSTAR} light curve (see Fig. \ref{fig:sliding-pprofs-nus}). Starting from the beginning of the 148~ks observation, we selected an initial segment of 20~ks and folded it at the best-determined spin period to generate a pulse profile. The time window was then shifted forward by 1~ks, and a new pulse profile was constructed using the subsequent 20~ks of data. This procedure was repeated iteratively until the end of the observation, producing a sequence of overlapping pulse profiles that track the evolution of pulse phase as a function of time. Each of the resulting profiles was stacked vertically as a function of pulse phase, with the horizontal axis representing the sliding time-window index. The same procedure was applied independently to the barycentered light curve (i) without and (ii) with a correction for the orbital motion with the \axsini\ determined from the \nustar\ data.

The left panel of Fig. \ref{fig:sliding-pprofs-nus} shows the systematic drift of the minima with time, appearing as an arc-like structure in phase. This is caused by the pulse arrival time delay due to the orbital motion. After correcting the light curves for orbital motion, the pulse minima appear to be aligned at a constant phase across all sliding windows. This explicitly shows that phase drift can be attributed to the orbital motion and is removed by the adopted orbital solution. A similar plot could not be generated from \rxte \ data due to its sparse sampling. 

\subsection{Spectral analysis} \label{sec:spec_analysis}

 In this work, we performed spectral modelling using \nustar\  observations only. The spectra from both FPMA and FPMB were optimally binned using the prescription suggested by \citet{Kaastra_2016} and were fit simultaneously while allowing for a freely varying cross-normalisation, $C_B$. This parameter was found to be around $C_B \sim$ 0.99. The spectral analysis was carried out in the 3-79 keV range using the package \textit{XSPEC} version:  12.13.1 \citep{Arnaud_1996}. The low-energy interstellar absorption was modelled using the \texttt{Tbabs} model \citep{Wilms_2000}, and the Fe $ K\alpha$ line was modelled using a Gaussian emission feature around 6.3~keV. 

 \begin{figure}
    \centering
    \vspace{0.5cm}
	\includegraphics[scale=0.40,trim={0cm 0cm 0 0},  angle=0]{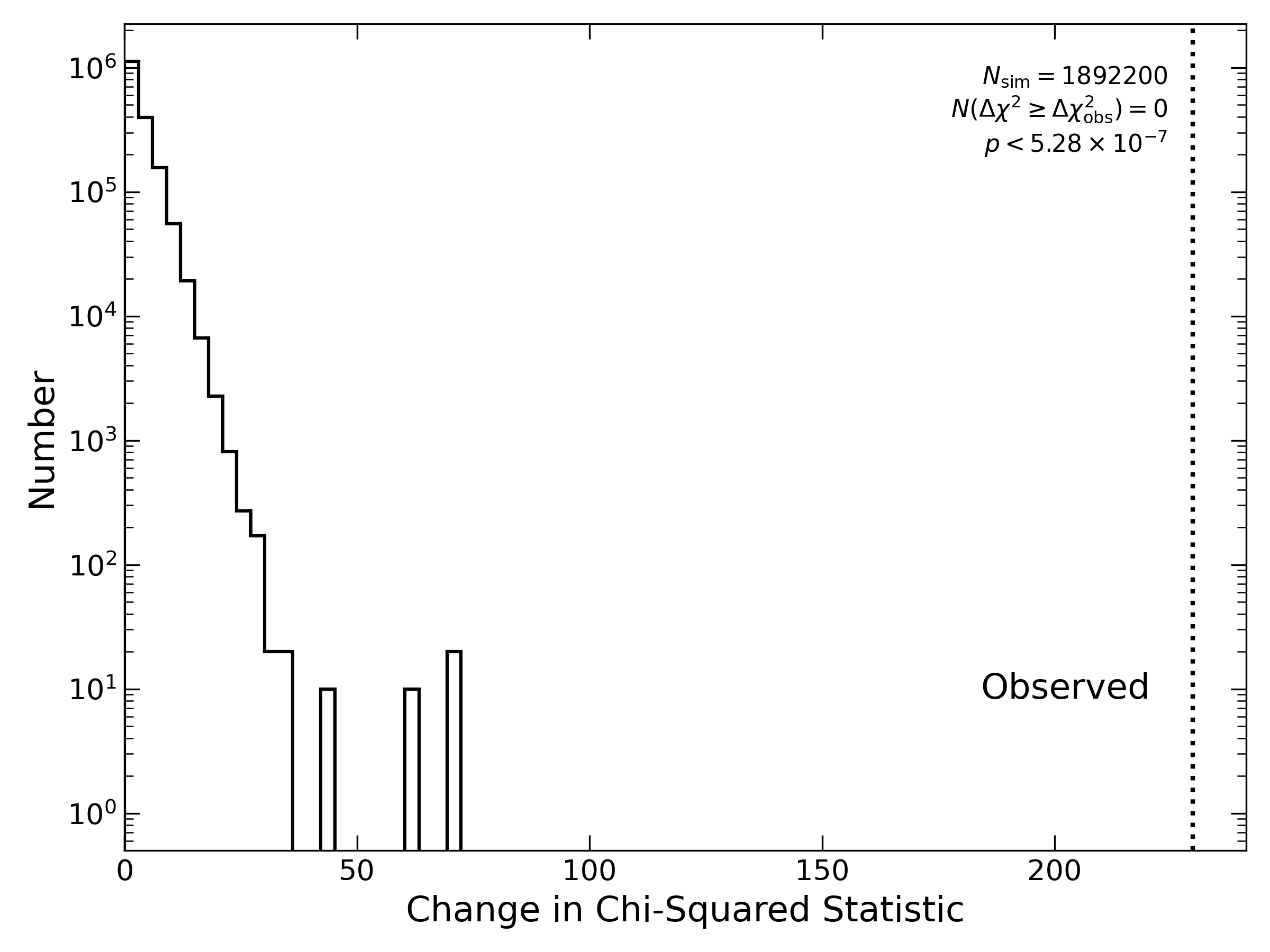}
    \caption{Distribution of the $\Delta\chi^2 $ values obtained from $N_{sim} = 1,892,200$ Monte Carlo simulations of the spectrum using \texttt{simftest}, comparing the fits with and without the cyclotron line under the \texttt{NPEX} continuum model. The dotted vertical line marks the $\Delta\chi^2 $ improvement obtained from the observed spectrum. No simulated realisations achieved a $\Delta\chi^2 $ equal to or exceeding the observed value, yielding a chance probability $p <5.3 \times 10^{-7}.$    }
    \label{fig:simftest_sig_test}
\end{figure}

To fit the continuum, we explored multiple models commonly used in modelling the spectra of accreting X-ray pulsars, such as \texttt{HighEC} \citep{White_1983}, \texttt{NPEX} \citep{Makishima_1999}, and the \texttt{CompTT} model \citep{Titrachuk_1994}. The results from different continuum models were comparable and have been presented in Table \ref{table:Bestfit-table-spec}. For brevity, we describe the fit with the \texttt{NPEX} model here, with the form 
\begin{equation}
\texttt{NPEX (E)}
=
A \left( E^{-\Gamma} + f\,E^{2} \right)
\exp\!\left( -\frac{E}{E_{\rm f}} \right)
\end{equation}

 Fitting the spectra with only the low‐energy absorption, continuum, and Fe K$\alpha$ line resulted in a poor fit ($\chi^2_{\text{red}} \sim$  1.5 for 421 d.o.f.) with pronounced, wave‐like residuals near 50~keV (Fig. \ref{fig:nustar-spectrum}a). Adding a Gaussian‐shaped absorption component (\texttt{gabs}) defined as
\begin{equation}
\texttt{GABS (E)}=\mathrm{exp}\Bigg\{ -\tau \cdot \mathrm{exp}\Bigg[-\frac{(E-E_\mathrm{0})^2}{2\sigma^2} \Bigg]\Bigg\}\\
\end{equation}
 at  $\sim 50$~keV (the cyclotron feature) yields a statistically acceptable solution ($\chi^2_{\text{red}} \sim$1 for 418 d.o.f.). In \textit{XSPEC} notation the final model is “\texttt{TBabs} $\times$ (\texttt{NPEX} + \texttt{gauss})\ $\times$ \texttt{gabs}$_{\text{cyc}}$”. Fig. \ref{fig:nustar-spectrum}b shows the resulting residuals after including the cyclotron line. The requirement of a \texttt{gabs} component for a good fit across the different continuum choices in Table~\ref{table:Bestfit-table-spec} shows that the cyclotron line is not an artefact of the continuum model.

To verify the robustness of the cyclotron absorption feature, we fit the spectrum with several different continuum models.  Each model yielded statistically equivalent fits and consistently recovered a cyclotron line at $\sim50$~keV (see Table \ref{table:Bestfit-table-spec}). The standard F-test routine in \textit{XSPEC} is unsuitable for evaluating the significance of nested components in a model. Therefore, to quantify the line's significance, we computed the probability of chance improvement (PCI) from the ratio of variances, which is not subject to this limitation, using the \textit{mpftest} routine\footnote{\url{https://www.nv5geospatialsoftware.com/docs/mpftest.html}} as suggested by \citet[see their Appendix A]{Orlandini_2012} and used in \citet{Devaraj_2025}. In every continuum model, the PCI was found to be below $6.5 \times10^{-5}$. To further assess the significance of the cyclotron line, we performed $\sim$1.89 million Monte Carlo simulations of the spectrum using \texttt{simftest} routine, comparing the fits with and without the cyclotron line assuming the \texttt{NPEX} continuum model. In none of the simulated realisations did the $\Delta\chi^2$ reach or exceed the value in that of the observed spectrum ($\Delta \chi^2 \approx 230$; see Fig. \ref{fig:simftest_sig_test}), placing an upper limit on the chance probability, $p < 5.3 \times 10^{-7}$, consistent with our earlier PCI estimate. This shows the high statistical significance of the detected cyclotron line feature.

\begin{table*}
 \bgroup
 \def\arraystretch{1.9}
 \caption{Best-fitting phase-averaged spectral parameters. Errors are reported at 90 \% confidence. The spectral analysis is discussed in Section \ref{sec:spec_analysis}}
\begin{center}
\begin{tabular}{|l|c|c|c|}
\hline 
\hline
Parameter                                           & HighEC                            & NPEX                               & CompTT                           \\
\hline
\hline
C$_{B}$ (constant factor)                           & $0.997 \pm 0.005$      & $0.997 \pm 0.005$       & $0.997 \pm 0.005$      \\
N$_{\mathrm{H}}$ [$10^{22}\,\mathrm{cm}^{-2}$]      & $5.4 \pm 0.3$            & $3.94 \pm 0.35$             & $1.3 \pm 0.5$           \\
Photon Index $\Gamma$                             & $1.38_{-0.03}^{+0.02}$         & $0.68 \pm 0.03$          & \ldots                           \\
E$_{\mathrm{cut}}$ (keV)                            & $33.1_{-0.9}^{+1.2}$            & \ldots                             & \ldots                           \\
E$_{\mathrm{fold}}$ (keV)                           & $15.2_{-3.5}^{+6.3}$              & $7.86_{-0.38}^{+0.64}$                             & \ldots                           \\
f$_{\mathrm{NPEX}}$ [\(\times10^{-4}\)]             & \ldots                            & $5.3_{-1.2}^{+0.9}$              & \ldots                           \\
Continuum Norm [\(\times 10^{-2}\)]$^*$                  & $1.49_{-0.09}^{+0.08}$            & $0.82_{-0.05}^{+0.06}$             & $0.25 \pm 0.01$           \\
CompTT $T_0$ (keV)                                  & \ldots                            & \ldots                             & $1.08 \pm 0.04$        \\
CompTT $kT$ (keV)                                   & \ldots                            & \ldots                             & $8.3 \pm 0.3$           \\
CompTT $\tau$                                       & \ldots                            & \ldots                             & $4.8 \pm 0.1$           \\
E$_{\mathrm{Fe}}$ (keV)                             & $6.28 \pm 0.06$         & $6.27 \pm 0.06$          & $6.29 \pm 0.06$        \\
$\sigma_{\mathrm{Fe}}$ (keV)                        & $0.16 \pm 0.10$         & $0.18 \pm 0.09$          & $0.22_{-0.08}^{+0.09}$        \\
Norm$_{\mathrm{Fe}}$ [\(\times 10^{-5}\)]$^{\dagger}$            & $5.6_{-1.4}^{+1.6}$               & $5.8_{-1.4}^{+1.6}$                & $6.7_{-1.5}^{+1.7}$              \\
Fe Eq width (eV)   &   $48\pm 13$       &   $51\pm 14$      & $63\pm 15$ \\ 
E$_{\mathrm{cyc}}$ (keV)                            & $52.1 \pm 4.6$              & $49.9_{-1.5}^{+2.0}$               & $49.8_{-1.3}^{+1.6}$             \\
$\sigma_{\mathrm{cyc}}$ (keV)                       & $13.6_{-1.6}^{+2.2}$              & $8.0_{-1.2}^{+1.6}$                & $7.56_{-0.9}^{+1.1}$            \\
$\tau_{\mathrm{cyc}}$                               & $1.17_{-0.35}^{+0.48}$            & $1.17_{-0.19}^{+0.26}$             & $1.16_{-0.15}^{+0.17}$           \\
\hline
Flux (1–100 keV) [\(\times10^{-10}\)]$^{\ddagger}$   & $2.66 \pm 0.05$          & $2.67 \pm 0.03$          & $2.67 \pm 0.02$         \\
Tot‐$\chi^2$ (d.o.f.) [w/ cyc]                       & $419.65\,(415)$                   & $398.04\,(418)$                    & $403.00\,(415)$                  \\
Tot‐$\chi^2$ (d.o.f.) [w/o cyc]                      & $615.82\,(418)$                   & $627.64\,(421)$                    & $706.50\,(418)$                  \\

PCI                                                 & $6.5\times10^{-5}$               & $2.5\times10^{-6}$                 & $9.9\times10^{-9}$               \\
\hline
\end{tabular}
 \end{center}
  \label{table:Bestfit-table-spec}

 \egroup
\begin{tablenotes}
 \item[*] $^*$In units of photons keV$^{-1}$cm$^{-2}$s$^{-1}$
 \item[*] $^\dagger$In units of photons cm$^{-2}$s$^{-1}$
 \item[*] $^{\ddagger}$Absorbed flux in units of \funi 
\end{tablenotes}
\end{table*}

\section{Discussion and Conclusion}

The discovery of SFXTs was made over two decades ago, and yet the mechanism driving the rapid X-ray variability has yet to be successfully explained \citep{Sidoli_2017}. Some of the competing models include the framework of the clumpy wind model to explain the highly variable behaviour. The winds from the companion are expected to contain inhomogeneous dense clumps of matter embedded in a lower density inter-clump medium \citep{Ducci_2009}. When the compact object in orbit captures these dense clumps, sharp bright X-ray flares are observed. However, when the rarer low-density clumps are accreted, the source luminosity is found to be much lower. From hydrodynamical simulations of line-driven wind from the companion, \citet{Oskinova_2012} find that the model predicts variability of up to eight orders of magnitude, for which there is a lack of observational evidence. The suppression of the accretion rate is necessary to explain the long periods of quiescence followed by the rapid, bright flaring activity \citep{Sidoli_2017}. 

This inhibition of accretion can also be explained if the compact object is a highly magnetised neutron star. When the mass flux is low (accretion of matter from the inter-clump medium), the magnetospheric radius can expand beyond the co-rotation radius and result in the inhibition of accretion due to the \textit{propeller effect} \citep{Bozzo_2016}, thereby reducing the X-ray luminosity to levels consistent with quiescence. When large clumps of wind are accreted, the ram pressure is sufficiently large, and the matter can penetrate the magnetosphere of the compact object, with the resulting accretion producing bright X-ray flares.

The detection of pulsations in several SFXTs provides compelling evidence that these systems are more likely to host neutron stars rather than black holes. This conclusion is further bolstered by the detection of cyclotron lines in these systems. The first detection of a cyclotron line in an SFXT was reported by \citet{Bhalerao_2015} in IGR J17544--2619. However, subsequent observations of the same source did not show the presence of this line \citep{Bozzo_2016}, underscoring the tentative and often transient nature of cyclotron line detections in these systems. A possible cyclotron line around 3.3 keV in IGR J18483--0311 was discussed by \citet{Sguera_2010}; however, this is yet to be confirmed.  

IGR J16418--4532 was proposed to be a candidate cyclotron line source with the line centred around 50~keV \citep{Natalucci_2024}. As discussed in section \ref{sec:spec_analysis}, we confirm the robust detection of the line in this work, indicating a magnetic field strength of $\sim 5 \times 10^{12}$~G. Cyclotron lines serve as the best diagnostic probes of magnetic fields in accreting pulsars. Their correlation with luminosity is linked to accretion regime transitions \citep{Becker_2012}. Observing these correlations enables a deeper understanding of the accretion column geometry and the underlying X-ray emission processes. Several cyclotron line sources, such as GX 304--1 \citep{Rothschild_2017} and Cep X--4 \citep{Vybornov_2017}, show only one type of correlation with luminosity, while those (such as A0535+26, \citealt{Shui_2024}) observed over a large dynamic range show a change in the nature of the correlations, indicating accretion regime transitions. Such a study involving the luminosity dependence of the cyclotron line in IGR J16418--4532 may be performed in the future by observing the source across a range of X-ray flux/luminosity states.

The masses of the companions in SgXBs are determined through the spectral classification of the donor's Optical/NIR spectra, which help identify the spectral type and luminosity class. This, in turn, yields stellar parameters such as effective temperature, luminosity, and surface gravity through atmosphere model calibrations \citep{Martins_2005}. By identifying the star's position on the HR diagram's evolutionary tracks, one can infer the stellar mass. The method is limited by the degeneracy in classifying spectral types. In the case of IGR J16418--4532, the nature of the donor remains somewhat uncertain. SED modelling of the optical/NIR spectra favors a BN0.5 Ia supergiant primarily due to the presence of strong He I absorption features \citep{Coleiro_2013} while the constraints derived from the system's eclipse duration and the requirement that the donor must fill its Roche lobe indicate the presence of an early O-type supergiant, with \citet{Coley_2015} arguing for a spectral type close to O7.5 I. 

The standard method for determining the orbital parameters involves measuring the phase shift of the pulse profiles and requires a stable pulse profile. However, as shown in Fig. \ref{fig:sliding-pprofs-nus}, the pulse profiles exhibit significant changes within a single observation. Yet a clear dependence of the pulsation detection on the \axsini \ can be seen in Fig. \ref{fig:efs-axsini-period-nus-rxte}. In this work, we corrected the photon arrival times for a range of trial \axsini \ values and found that there is a marked improvement in the significance of the pulsation %
at \axsini\ around 55 lt-s (assuming a circular orbit).

\begin{figure}[t]
    \centering
    \includegraphics[scale=0.43, angle=0]{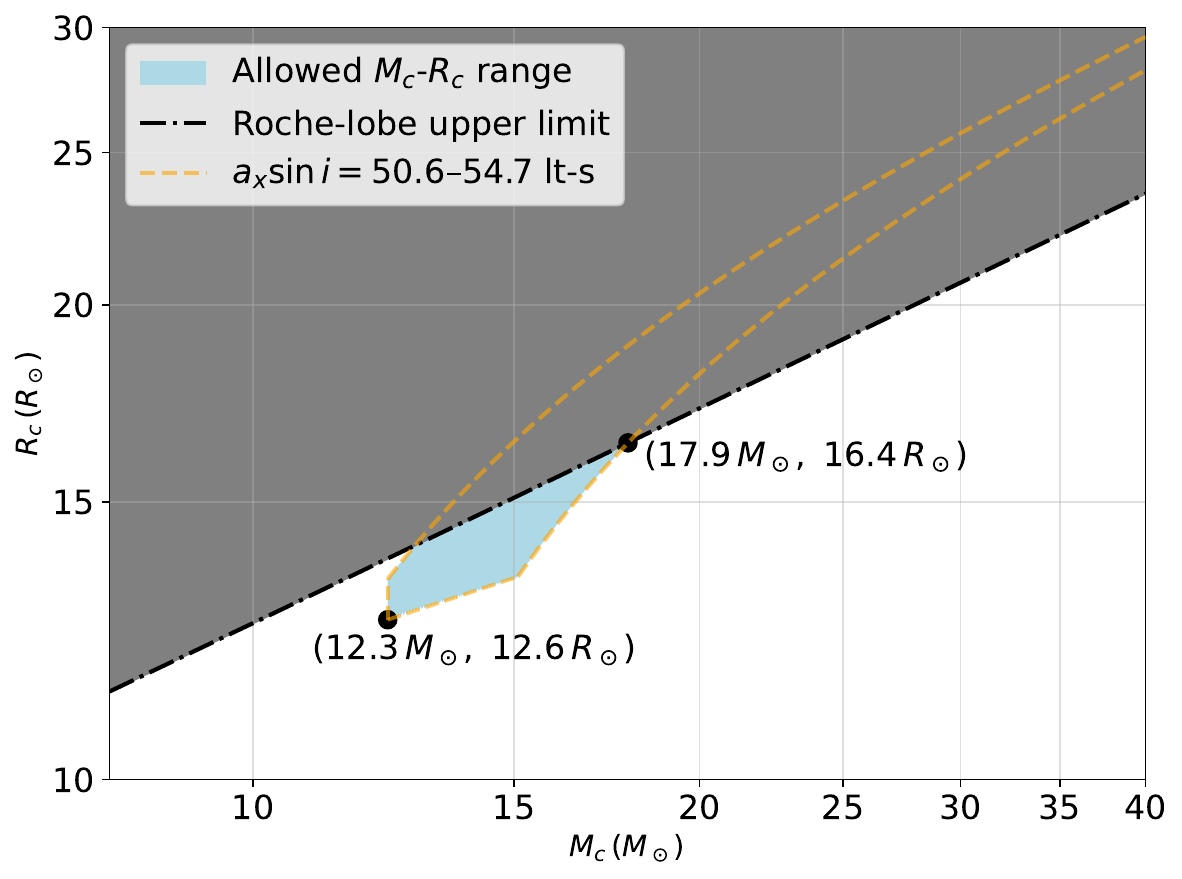}
    \caption{Mass-radius constraints on the companion star from eclipse geometry and Roche-lobe considerations. The dashed orange curves show the allowed range of companion mass and radius within the $1\mathrm{-}\sigma$ confidence level, obtained from the eclipse geometry using the weighted mean of the \axsini\ values  ($50.6\text{--}54.7$~lt-s) from the \textit{RXTE} and \textit{NuSTAR} timing analyses. The Roche-lobe upper limit is indicated by the dash-dotted black curve, with forbidden radii shaded in grey. The blue shaded region represents the companion mass-radius parameter space that simultaneously satisfies the eclipse-geometry constraint and the Roche-lobe constraint. The overlap of these constraints defines the physically allowed parameter space, with the black points marking the lower and upper limits on the companion
mass and radius.}
    \label{fig:mc_rc_constraints}
\end{figure}

The orbital dynamics of the system is governed by Kepler’s third law, which relates the orbital separation $a$ to the orbital period $P_{\rm orb}$ and the total system mass,
\begin{equation}
a = \left[\frac{G\,(M_{\rm x}+M_{\rm c})\,P_{\rm orb}^2}{4\pi^2}\right]^{1/3},
\end{equation}
where $M_{\rm x}$ and $M_{\rm c}$ are the masses of the neutron star and companion, respectively, and $G$ is the gravitational constant. The neutron star mass is assumed to be $1.4\,M_\odot$ in this analysis. The neutron star orbits the centre of mass with a semi-major axis
$a_{\rm x} = a\,{M_{\rm c}}/({M_{\rm x}+M_{\rm c}})$.
Pulse-timing analysis of the \textit{RXTE} and \textit{NuSTAR} data constrains the projected semi-major axis \axsini, which, for a given choice of component masses, fixes the orbital inclination $i$.

To further restrict the system parameters, we incorporate geometric constraints from the eclipse duration and Roche-lobe considerations \citep{Hill2005, Islam2015}. The companion star's radius is then obtained from the eclipse geometry \citep{Rubin_1996},
\begin{equation}
R_{\rm c} = a\left(\cos^2 i + \sin^2 i ~\sin^2\theta_E\right)^{1/2}.
\end{equation}
This procedure yields an eclipse-based mass–radius relation for the companion star. Since the two \axsini \ measurements agree within $1\sigma$, we calculate their inverse-variance weighted mean, \axsini \ = $52.7\pm 2.1 $ lt-s, and combine this eclipse-based relation with it to determine the physically allowed region in the companion mass–radius plane. This region, satisfying both the eclipse-geometry constraint (dashed orange curves, marking the $1\sigma$ bounds \axsini \ = $50.6 - 54.7$~lt-s) and the Roche-lobe upper limit (dash-dotted curve), is shown as the blue region in Fig.~\ref{fig:mc_rc_constraints}.

An additional upper bound on the companion radius is imposed by the Roche-lobe criterion \citep{Bowers_1984, Hill2005},
\begin{equation}
R_{\rm Roche} = a \left( 0.38 + 0.2 \log_{10} \frac{M_{\rm c}}{M_{\rm NS}} \right),
\end{equation}
which limits the maximum extent of the companion star for a given mass. This constraint is shown in Fig.~\ref{fig:mc_rc_constraints} as the dash-dotted black curve, with the shaded grey region indicating Roche-lobe overflow and thus a forbidden parameter space. The overlap of the eclipse-based mass-radius constraints and the Roche-lobe constraint is highlighted by the blue shaded region in Fig.~\ref{fig:mc_rc_constraints}. This region defines the physically allowed parameter space and constrains the companion star to have a mass in the range $\simeq$12$-$18~$M_\odot$ and a radius in the range $\simeq$13$-$16 $R_\odot$.

We would like to point out that even a small variation of the pulse profile with time, or with orbital phase, can add unknown systematic errors to the \axsini \ \citep{Raichur_2010}. The light curves of the SFXTs include many flares, with timescales comparable to the pulse period, resulting in variations in the pulse profile with time. As the light travel time across the binary is a small fraction of the pulse period, small variations in the pulse profiles as seen in Fig. \ref{fig:sliding-pprofs-nus} would lead to systematic errors in the formal pulse arrival time calculations. In the current work, we have optimised the total pulse detection with trial values for the size of the binary. However, there are unknown systematics due to the recurring rapid flares and intrinsic pulse profile variations, if any. Still, the overall consistency between the \textit{NuSTAR} and \textit{RXTE/PCA} results is reassuring. 

Dynamical measurement constraints seem to favour a substantially lower companion radius and mass, $R_c \simeq$13$-$16 $R_\odot$ and $M_c \simeq12-18M_\odot$, than the $R_c \gtrsim 20.8R_\odot$ and $M_c \gtrsim 36 M_\odot$ inferred by \citet{Coley_2015}. The two estimates are arrived at using different approaches and hence different quantities. Our constraints depend on the measured projected semi-major axis values, for which $f(M) \propto (a_x~sin~i)^3$. Systematic uncertainties in $a_x~sin~i$ are amplified in the inferred mass as mentioned earlier. \citet{Coley_2015} combined Roche-lobe constraints with spectral-type dependent stellar masses and radii adopted from O-type star calibrations \citep{Martins_2005}. Their lower limits are established based on the earliest spectral type to have radii restricted within the Roche lobe, rather than a direct dynamical measurement. Thus, the discrepancy likely reflects the different systematic sensitivities of the two approaches; notably, the disagreement is much stronger in mass than in radius, with the latter differing only at the $\sim 20\%$ level. 

\section{Summary}

In this paper, we have reported the results from the analyses of data from \nustar\ and \rxte\ observations of the intermediate SFXT, IGR J16418--4532. Coherent pulsations of $P_{\mathrm{spin}} = 1208.67 \pm 0.04$~s and $P_{\mathrm{spin}} = 1209.4178 \pm 0.0084$~s have been detected in both the \nustar\ and \rxte\ observations, respectively. For the first time, we have estimated the binary mass function of an SFXT using the pulse time delay. The time of arrival of photons was corrected for a range of \axsini \ values and found that the detection significance of the pulsations showed clear dependence on the choice of \axsini. Using the weighted mean of \axsini \ values from the two timing analyses,  together with the eclipse-geometry and Roche-lobe constraints, we constrain the companion star to have a mass in the range $\sim$12$-$18 $M_\odot$ and a radius in the range $\sim$13$-$16 $R_\odot$. We successfully modelled the spectral continuum using \texttt{HighEC}, \texttt{NPEX}, and \texttt{CompTT} models with comparable $\chi^2$. A cyclotron line is detected independent of the choice of continuum models around 50~keV, confirming its status as a cyclotron line source.

\begin{acknowledgements}
AD would like to thank Varun Bhalerao for his useful comments that improved the quality of this manuscript. AD acknowledges the CSR support of BDP UGL Global Logistics Pvt. Ltd., which enabled access to the computational resources used in this study. We thank the referee for the constructive feedback that has helped improve the clarity of this manuscript.
\end{acknowledgements}

\bibliographystyle{aa_url} %
\bibliography{bibtex}

\end{document}